\documentclass{cambridge6A}
   
\usepackage{amsmath,amssymb}
\usepackage{graphicx}
\usepackage{verbatim}

\newcommand{\be}{\begin{equation}}
\newcommand{\ee}{\end{equation}}
\newcommand{\bea}{\begin{eqnarray}}
\newcommand{\eea}{\end{eqnarray}}

\begin{document}

\renewcommand*\thesection{\arabic{section}}
\renewcommand*\thesubsection{\arabic{section}.\arabic{subsection}}

\setcounter{equation}{0}
\numberwithin{equation}{section}
\setcounter{figure}{0}
\renewcommand{\thefigure}{\arabic{figure}.}

\chapterauthor{Gary T. Horowitz$^{1}$ and Toby Wiseman$^{2}$\\ ~ \\
\textit{${}^{1}$
Department of Physics, UCSB, Santa Barbara, CA 93106\\${}^{2}$Blackett Laboratory, Imperial College, London, SW7 2AZ} }

\chapter*{General black holes in Kaluza-Klein theory}

One of the oldest ideas for unifying gravity and electromagnetism is to consider general relativity
in five dimensions with one dimension curled up into a small circle.  In other words, one studies vacuum solutions in five dimensions which asymptotically approach ${\mathbb R}^4\times S^1$ at infinity. This is known as Kaluza-Klein theory \cite{Kaluza:1921tu,Klein:1926tv}. We have already discussed the simplest black hole solution in this theory: the product of Schwarzschild and a circle. In this chapter we discuss more general Kaluza-Klein black holes. We begin by describing a surprising property of the total energy in this theory. We then discuss black holes which are invariant under translations around the circle, and then drop this restriction.

\copyrightline{Chapter of the book \textit{Black Holes in Higher Dimensions} to
be published by Cambridge University Press (editor: G. Horowitz)}

\section{Energy in Kaluza-Klein theory}

 The total energy of an asymptotically flat five dimensional spacetime  is given by a direct generalization of the standard ADM formula in four dimensions. Consider the metric on an asymptotically flat spacelike surface and let $x^i$ be asymptotically Euclidean coordinates. Then  
 \be
 g_{ij} =\delta_{ij} + h_{ij}
 \ee
and the total mass is given by: \index{mass}
\be\label{kkmass4}
M= {1\over 16\pi G_5} \oint (h_{ij,j} - h_{jj,i})dS^i
\ee
where the integral is over a surface at large $r$. In Kaluza-Klein theory, an identical formula holds, but the integral is now over a surface with topology $S^2 \times S^1$, where the $S^2$ has radius $r$ and the $S^1$ has length $L$. The  Kaluza-Klein vacuum, the product of a circle and four dimensional Minkowski space, has $M = 0$.

With standard asymptotically flat boundary conditions, there is a positive energy theorem which states, roughly speaking, that the solution with lowest total energy is Minkowski spacetime. More precisely, recall that the dominant energy condition \index{energy condition!dominant} states that $T_{\mu\nu} t^\mu_1 t_2^\nu \ge 0$ for any two future directed timelike vectors $t_1^\mu, t_2^\nu$.  One can prove the following \cite{Schoen:1979zz,Witten:1981mf}: \index{positive energy theorem}
\vskip .3cm
{\bf Positive energy theorem:} Consider any nonsingular, asymptotically flat initial data set satisfying the dominant energy condition. Its total energy  cannot be negative and vanishes only for flat Minkowski spacetime. 
\vskip .3cm
  The condition that the initial data be nonsingular is required to rule out, e.g., the negative mass Schwarzschild solution. This result holds in all dimensions $D\ge 4$ \cite{Dai:2004yd}, but only if the spacetimes asymptotically approach $\mathbb{R}^D$ with the flat Minkowski metric. 
  
  If a five dimensional spacetime  asymptotically approaches $\mathbb{R}^4\times S^1$ at infinity as required in  Kaluza-Klein theory, then this theorem does not hold \cite{Witten:1981gj}. To see this, consider the five dimensional generalization of the Schwarzschild metric:
 \be\label{schw5d4}
ds^2 = - \left(1-\frac{r_0^2}{r^2}\right)\  d t^2 + \left(1-\frac{r_0^2}{r^2}\right)^{-1}{dr^2} + r^2(d\theta^2 + \sin^2\theta d\Omega^2 ) \; .
\ee
We now analytically continue $t\to i \chi$ and $\theta \to \pi/2 + i\tau$. The result is
 \be\label{bubble4}
ds^2 = \left(1-\frac{r_0^2}{r^2}\right)\  d \chi^2 + \left(1-\frac{r_0^2}{r^2}\right)^{-1}{dr^2} + r^2(-d\tau^2 + \cosh^2\tau d\Omega^2 ) \; .
\ee
Since (\ref{schw5d4}) is Ricci flat, so is (\ref{bubble4}). To investigate the behavior near $r = r_0$, set $r = r_0 + \rho^2/2r_0$. Then near $\rho =0$, the $(\chi,r)$ part of the metric becomes
\be
d\rho^2 + \rho^2 {d\chi^2 \over r_0^2} \; .
\ee
So if $\chi$ is periodic with period $2\pi r_0$, then $\rho = 0 $ ($r=r_0$) is just a smooth origin of a rotational symmetry. Now consider the metric on the $\tau =0$ surface.  This spatial metric is asymptotically flat, and since $\chi$ is periodic, it satisfies Kaluza-Klein boundary conditions. What is the total energy of this solution? From (\ref{kkmass4}) the energy comes from the $1/r$ correction to the metric, but since there is no $1/r$ correction, the total energy vanishes! This shows that $M_4\times S^1$ is not the only zero energy solution in the theory. 

What is the interpretation of the solution (\ref{bubble4})? Since the $\chi$ circle smoothly caps off at $r=  r_0$, there is no spacetime for $r < r_0$. This is called a ``bubble of nothing", in analogy to a bubble of true vacuum which can nucleate inside a false vacuum in field theory. \index{bubble of nothing} This ``bubble of nothing" expands rapidly until it is moving close to the speed of light and eventually hits null infinity.

Not only are there nontrivial solutions with zero energy, one can find vacuum solutions with arbitrarily negative energy \cite{Brill:1989di,Brill:1991qe}. Consider time symmetric initial data, i.e., initial data with zero extrinsic curvature just like the $\tau = 0 $ surface above. The only constraint on the spatial metric is that the scalar curvature vanishes. We now use the fact that the  Reissner-Nordstr\"om metric has zero scalar curvature since the stress tensor for a Maxwell field in four dimensions is traceless. Analytically continuing $t\to i\chi$ and $q \to iq$ in the Reissner-Nordstr\"om metric yields
\be
ds^2 = U(r) d\chi^2 + U^{-1}(r)  dr^2 + r^2 d\Omega^2
\ee
with
\be
U(r) = 1- {2m\over r} - {q^2\over r^2} \; .
\ee
$U(r)$ vanishes when $r=r_+ \equiv m + \sqrt{m^2 +q^2} $, so we restrict
the $r$ coordinate to $r\ge r_+$. To avoid a conical singularity
at $r=r_+$, we identify $\chi$ with period
\be
     L = \frac{4\pi} {U'(r_+)}   = {2\pi r_+^2\over r_+ - m} \; .
     \ee
(Note that we are interested only in the metric and not the
Maxwell field that is also a part of the usual Reissner-Nordstr\"om solution.
We are constructing initial data for a five dimensional vacuum solution.
 In particular, $q$ should not be thought of as an electromagnetic
charge.)
     
One can easily evaluate the energy (\ref{kkmass4}) for this initial data and find $G_5 M = mL/2$. The parameter $m$ can be positive or negative in this construction since $r_+ > 0$ and the initial data remains nonsingular. Thus we have negative energy solutions.
(The special case $m=0$ is precisely the initial data for the ``bubble of nothing" solution discussed above.) By taking $m\to -\infty$ and $q \approx (-m)^{3/4}$, one can make $M$ arbitrarily negative keeping $L$ constant.

Given that there are solutions with arbitrarily negative energy in Kaluza-Klein theory, one might wonder why it is still taken seriously as a physical theory. The main reason is that all of the solutions we have discussed which violate the positive energy theorem have the property that spinors must be antiperiodic around the circle at infinity. This is a problem for the following reason: One can expand any five dimensional fermion in a Fourier series around the $S^1$. The result is an infinite tower of four dimensional fields with masses which are multiples of $1/L$. Since we want $L$ to be small, the only light fermion is the zero mode. In a space in which all femions are antiperiodic,  there is no such zero mode. Hence all fermions will be very massive which contradicts the fact that we observe light fermions in nature. A realistic theory requires that we  supplement the boundary condition with the requirement that fermions are periodic around the circle at infinity. With this added condition, one can prove a positive energy theorem \cite{Dai:2004yd}. All of the solutions we discuss in the remainder of this chapter will satisfy this condition.

\section{Homogeneous black hole solutions}


If a five dimensional spacetime is invariant under translations around a small circle, it can be viewed as an effective four dimensional spacetime coupled to certain matter fields. We begin by constructing the four dimensional theory which governs these fields. Letting $y$ be a periodic coordinate around the circle with period $L$, we can write the five dimensional metric
 in the form
\be\label{kkmetric4}
{ds^2=e^{-4\phi/\sqrt{3}} (dy+2A_\mu dx^\mu)^2 + e^{2\phi/\sqrt{3}} 
g_{\mu\nu} dx^\mu dx^\nu}
\ee
where $\phi, A_\mu, g_{\mu\nu}$ depend on $x^\mu$ but not $y$. The reason for the strange factor of $\sqrt 3$ in the definition of the scalar $\phi$ will become clear shortly.

Note that under a coordinate transformation, $y=\tilde y + 2\lambda (x^\mu)$, 
\be
dy + 2A_\mu dx^\mu=d\tilde y + 2\tilde A_\mu dx^\mu
\ee
 where $\tilde A_\mu = A_\mu + \partial
_\mu \lambda$. So $A_\mu$ is like a four dimensional Maxwell potential and part of the usual five dimensional coordinate transformations
are ordinary gauge transformations for $A_\mu$.

 For the metric (\ref{kkmetric4}),  $\sqrt{-^5g} = e^{2\phi/\sqrt{3}} \sqrt{-g}$
(independent of $A_\mu$). If we evaluate the Einstein action
for metrics of this form, we find (dropping a surface term):
\begin{eqnarray}
S & = & \frac{1}{16\pi G_5} \int dy\, d^4x\ \sqrt{-^5g}\ ^5R \nonumber\\
& = & \frac{1}{16\pi G_4} \int d^4x\ \sqrt{-g}\ \left[R-2(\nabla\phi)^2 -e^{-2\sqrt{3}
\, \phi} F^2\right]
\end{eqnarray}
where $F_{\mu\nu}\equiv 2\nabla_{[\mu} A_{\nu]}$ is the Maxwell field associated
with $A_\mu$, and $G_4=G_5/L$.  

Thus, for metrics which are independent of $y$,  the five dimensional Einstein action reduces to the action for  four dimensional general relativity, a Maxwell field, and an extra scalar
field $\phi$. The strange factors in (\ref{kkmetric4}) were chosen so that $g_{\mu\nu}$ has the standard four dimensional Einstein action. 
The equations of motion which follow from this action are
\be
\nabla_\mu \left(e^{-2\sqrt{3}\, \phi}\, F^{\mu\nu}\right)=0
\ee
\be
\nabla^2\phi + \frac{\sqrt{3}}{2}\ e^{-2\sqrt{3}\, \phi}\, F^2=0
\ee
$$G_{\mu\nu}  =  2\nabla_\mu \phi\, \nabla_\nu \phi - g_{\mu\nu}\, (\nabla \phi)^2$$
\be
+ 2e^{-2\sqrt{3}\, \phi}\, F_{\mu\rho} F_\nu^\rho -\frac{1}{2}\ g_{\mu\nu}
\, e^{-2\sqrt{3}\, \phi}\, F^2 \; .
\ee
The five dimensional mass, computed from (\ref{kkmass4}), agrees with the standard mass computed from the four dimensional metric $g_{\mu\nu}$.

One simple class of solutions to these equations is  $\phi=$ constant, $F_{\mu\nu}=0$, and $g_{\mu\nu}$ is any four dimensional vacuum solution of Einstein's
equation. This is just saying that if a four dimensional metric, $ds^2_4$, is Ricci flat, then its product with
$S^1$
\be
ds^2_5=ds^2_4 + dy^2
\ee
 is also Ricci flat. In particular, if $ds^2_4$ is the Schwarzschild metric, we get the black string discussed in the previous two chapters. 
 
 The above relation between the four and five dimensional Newton's constant helps explain a puzzle about black hole entropy. \index{black hole thermodynamics!entropy} In five dimensions, the entropy is proportional to the three dimensional ``area" of the event horizon. Since this clearly increases with the length of the circle, one might think that the five dimensional black string can have many more microstates than the four dimensional black hole. But in fact the five dimensional entropy is the same as the four dimensional entropy:
 \be
 S_5 = {A_5\over 4G_5} = {A_4 L\over 4G_5} = {A_4 \over 4G_4} =S_4 \; .
 \ee


\subsection{Nonrotating charged black holes}

We have seen that neutral black holes can be obtained by simply taking a product of the Schwarzschild solution and a circle. The same is not true for charged black holes. These are not simply related to the Reissner-Nordstr\"om solution, since if $F_{\mu\nu} \not=0$, it acts as
a source for $\phi $ and $ \phi$ cannot be constant.  At first sight it seems difficult to find charged black hole solutions in this theory since the field equations involve exponentials of $\phi$. However one can use a trick. From the five dimensional standpoint, one can generate a Maxwell field by a simple boost in the $y$ direction. Indeed, the four dimensional charge is simply the five dimensional momentum. So we start with the product of Schwarzschild and a line:
\be
ds^2 = - \left(1-\frac{2m}{r}\right)\  d\hat t^2 + \left(1-\frac{2m}{r}\right)^{-1}{dr^2} +
r^2d\Omega^2 + d\hat y^2 \; .
\ee
We now boost this solution in the $\hat y$ direction
\begin{eqnarray}
\hat t & = & t\, \cosh\alpha - y\, \sinh\alpha\nonumber \\
\hat y & = & y\, \cosh\alpha - t\, \sinh\alpha\, .
\end{eqnarray}
Then\footnote{These boosted black strings are also subject to Gregory-Laflamme instabilities \cite{Hovdebo:2006jy}.},
$$
ds^2 = -\left(1-\frac{2m \cosh^2\alpha}{r}\right)\ dt^2 + 
\left(1+ \frac{2m\sinh^2\alpha}{r}\right)\ dy^2$$
\be\label{pureboost4}
- \frac{4m \cosh\alpha \sinh\alpha}{r}\ dt\, dy +
 \left(1-\frac{2m}{r}\right)^{-1}{dr^2} + r^2 d\Omega^2 \; .
\ee
We now compactify $y$ and
put this into standard Kaluza-Klein form by defining
\begin{eqnarray}
e^{-4\phi/\sqrt{3}} & = & 1 + \frac{2m\sinh^2\alpha}{r}\\
A_t & = & -\frac{m\cosh\alpha \sinh\alpha}{r+2m \sinh^2\alpha} \; .
\end{eqnarray}
In defining the four dimensional metric $g_{\mu\nu}$ one must remember to subtract the  $A^2_t$ term from $g_{tt}$.  Setting
\be
 q = 2m \cosh^2\alpha
 \ee
  the result is
\be\label{chargedkkbh4}
g_{\mu\nu} dx^\mu dx^\nu = {-fdt^2 + \frac{d r^2}{f} + R^2 d\Omega^2}
\ee
where 
\be\label{defR4}
f(r) = {r-2m\over [r^2+(q-2m)r]^{1/2}},  \qquad R^2(r) = r[r^2 +(q-2m)r]^{1/2} \; .
\ee

This is a black hole with event horizon at $r = 2m$ and singularity at $
r=0$. Note that the boost and dimensional reduction has not changed the location of the horizon or singularity. Unlike Reissner-Nordstr\"om, these black holes do {\it not} have a regular inner
horizon. The  singularity is spacelike, like the Schwarzschild solution. \index{singularity!spacelike}

The total mass can be obtained by applying the general formula (\ref{kkmass4}) to the five dimensional metric (\ref{pureboost4}), but an easier approach is to  compare the four dimensional metric (\ref{chargedkkbh4}) at large $r$ with the Schwarzschild solution. Expanding the metric functions for large $r$, we have $R^2 = r^2 + r(q-2m)/2 + \cdots$ and $f= 1-(q+2m)/2r+ \cdots$. To compare with Schwarzschild, we take $R$ to be our radial coordinate, so $f= 1-(q+2m)/2R = 1-2G_4 M/R$.  The total charge can be obtained by writing $A_t = -Q/R +   \cdots$. The results are
\be\label{defQ4}
G_4 M = {q+2m\over 4}, \qquad Q^2 = {q(q-2m)\over 4} \; .
\ee


The extremal limit corresponds to the maximum possible charge for a given mass. This limit corresponds to $m\rightarrow 0 $, so one obtains \index{extremal limit}
\be
Q\to \frac{q}{2}\, ,\quad G_4 M\to \frac{q}{4}\Rightarrow Q=2G_4 M \; .
\ee
Recall that the extremal Reissner-Nordstr\"om black hole has $Q=G_4 M$ so Kaluza-Klein black holes can carry twice as much charge. Actually, a Kaluza-Klein black hole cannot quite carry twice as much charge since the extremal limit is singular. The event horizon is at $r=2m$, so in the extremal limit  ($m=0$) the horizon becomes singular. Since the nonextremal black hole has a spacelike singularity, one might expect that in the extremal limit the singularity is either spacelike or null. However the following calculation shows that the singularity is actually timelike. 
 Consider a 
radial null geodesic that hits the singularity. It satisfies
\be
dt=\frac{dr}{f} \approx q^{\frac{1}{2}} \frac{dr}{r^{1/2}} 
\ee
where the last expression is valid for small $r$. 
Since the ingoing radial null geodesic can reach the singularity at $r=0$ in finite $t$, it can lie entirely to the past of an outgoing radial null geodesic.  This shows the singularity is timelike. \index{singularity!timelike} If the singularity were null, then $t$ would diverge as $r\to 0$, just like geodesics approaching a horizon.

What is the five dimensional description of an extreme black hole?  Writing (\ref{pureboost4}) in terms of $q = 2m\cosh^2\alpha$ and taking the limit $m\to 0$ keeping $q$ fixed yields
\be
ds^2={-dt^2+dy^2+dr^2+r^2d\Omega^2} + \frac{q}{r}\ (dt-dy)^2 \; .
\ee
This is a five dimensional generalization of a plane fronted gravitational wave. Setting $v= t+y$, $u = t-y$, plane fronted waves take the form 
\be
ds^2= -dudv + dx^idx_i  + F(u, x^i)\, du^2 \; .
\ee
These metrics are Ricci flat if $F$ satisfies a simple flat space Laplace equation:
\be
\partial^i \partial_i F=0 
\ee
where $i = 1,2, 3$ in five dimensions.
The $u$ dependence is arbitrary. The extremal black hole corresponds to a very simple plane fronted wave which  is independent of $u$. It is a higher dimensional version of the Aichelberg-Sexl metric. \index{Aichelberg-Sexl metric}

We have discussed electrically-charged solutions. Magnetically-charged solutions can be
obtained by a duality rotation: Setting
\be
^*F_{\mu\nu}=\frac{1}{2}\ e^{-2\sqrt{3}\phi} \ {\varepsilon_{\mu\nu}}^{\rho\sigma}\
F_{\rho\sigma}
\ee
 the field equations are invariant under $F\to {^*F}$, $\phi\to-\phi$. (A similiar duality rotation can be used to obtain magnetically charged Reissner-Nordstr\"om black holes.)
So a magnetically-charged black hole has the same  four dimensional metric (\ref{chargedkkbh4},\ref{defR4}) but different matter fields. To emphasize that the solution is now magnetically charged we replace $q$ with $p$, i.e., we set $p= 2m \cosh^2\alpha$. Then

\begin{eqnarray}
e^{4\phi/\sqrt{3}} & = & 1+\frac{p-2m}{r}\\
A_\phi & = & P(1-\cos\theta)\label{magpot4}
\end{eqnarray}
where
\be 
P^2 = {p(p-2m)\over 4} \; .
\ee
$P$ is  the magnetic charge since $F_{\theta\phi} = \partial_\theta A_\phi = P\sin\theta$, so
\be
P=\frac{1}{4\pi} \int F_{\theta\phi} d\theta\, d\phi \; .
\ee
There is no globally defined vector potential for a magnetic monopole, since if $A$ were globally defined, $\int_{S^2} dA =0$. The vector potential in (\ref{magpot4}) is not well behaved at $\theta = \pi$ since $\int A_\phi d\phi \ne 0$  even as the circle shrinks to zero size. This is sometimes called a ``Dirac string". To avoid it, one can work with two patches on the sphere, using (\ref{magpot4}) on the northern hemisphere and setting $A_\phi = -P(1+\cos\theta)$ on the southern hemisphere.

Let us reconstruct the five dimensional metric corresponding to the magnetically charged black hole. Using the general formula (\ref{kkmetric4})
we get
\begin{eqnarray}\label{magbs4}
ds^2 & = & \left(1+\frac{p-2m}{ r}\right)^{-1}\ \left[dy + 2P\, (1-\cos \theta)
\, d\phi\right]^2 \nonumber\\
 & -& {\left(1-{2m\over r}\right)}
\ d t^2 + \frac{\left(1+\frac{p-2m}{ r}\right)}{\left(1-\frac{2m}{ r}\right)}dr^2 +  r^2
\, \left(1+\frac{p-2m}{ r}\right)\, d\Omega^2 
\end{eqnarray}
Note that the square roots that had been present in the four dimensional metric are gone. This spacetime still has a horizon at $r=2m$ and singularity at $r=0$. We now take the
extremal limit: $m=0$. Remarkably,  $g_{ tt}=-1$, so the metric reduces to a simple product of time and a four dimensional positive
definite metric.  The four dimensional metric still looks singular at $r=0$, but if we set $\chi\equiv y/2P$ and
\be
\rho\equiv 2(pr)^{\frac{1}{2}} \Rightarrow 
d\rho^2 = \frac{p dr^2}{r}
\ee
 then near $r=0$ 
the spatial metric is
\be
d\rho^2 + \frac{\rho^2}{4}\ \Bigl\{[d\chi + (1-\cos\theta)\, d\phi]^2 + d\theta^2 +
\sin^2\theta\, d\phi^2\Bigr\} \; .
\ee
If $\chi$ is periodic with period $4\pi$, the quantity in brackets is the metric on $S^3$
with radius two expressed as a Hopf fibration. \index{Hopf fibration} So the metric is smooth at $\rho=0$ (or $r=0$) provided $y$ has period $8\pi P$. This  fixes the
period of the Kaluza-Klein circle in terms of the magnetic charge $P$. This nonsingular solution is called the {\it Kaluza-Klein monopole} \cite{Gross:1983hb,Sorkin:1983ns}.

It is a remarkable fact that Kaluza-Klein theory has a nonsingular magnetic monopole \index{magnetic monopole} solution since, e.g., Einstein-Maxwell theory does not\footnote{Of course the five dimensional metric is still a vacuum solution to Einstein's equation. The magnetic charge arises in the reduction to four dimensions.}. The spatial metric is called the Taub-NUT instanton  \index{Taub-NUT instanton} and has topology $\mathbb{R}^4$. Asymptotically, it looks like $\mathbb{R}^3\times S^1$ and satisfies the usual Kaluza-Klein boundary conditions, but at small $r$ the $S^1$ combines with the $S^2$ of spherical symmetry to form an $S^3$ which smoothly shrinks to zero at the origin of the four dimensional space.

Since the four dimensional metric is the same as the electrically charged case, the extremal limit is singular. Thus we have  an example of a four dimensional spacetime with curvature singularity that is resolved by lifting the solution to five dimensions. 
Conversely, one can see that the singularity in the four dimensional metric arises since $g_{yy} = 0$ at $r=0$. In other words, the length of the circle that we are reducing along goes to zero there. 

We have seen that the extremal limit of both the electrically charged and magnetically charged Kaluza-Klein black hole is singular in four dimensions. It turns out that black holes with both electric and magnetic charge (sometimes called dyonic black holes) have a nonsingular extremal limit with nonzero horizon area. We will not present it here, since the solution is considerably more complicated \cite{Gibbons:1985ac} and it is easily obtained from the more general rotating black holes discussed in the next section. 

\subsection {Rotating Kaluza-Klein black holes}

To obtain a rotating, electrically charged Kaluza-Klein solution, one can repeat the construction in the previous section starting with the Kerr metric. In other words, one takes the product of Kerr and a line, boosts along the line and then compactifies the extra dimension. Rather than discuss this solution explicitly,  we will jump ahead and give the most general known analytic family of black holes in Kaluza-Klein theory. These solutions are all stationary, axisymmetric, and invariant under translations in $y$. They depend on four parameters $(m,q,p,a)$ which determine the mass $M$, electric and magnetic charges $Q,P$ and angular momentum $J$. 
  The five dimensional metric is:

\be
ds^{2} = {H_{2}\over H_{1}}(dy + 2{\bf A})^{2} 
-{\Delta_\theta\over H_{2}}(dt + {\bf B})^{2}
+H_{1}\left({dr^{2}\over\Delta}+d\theta^{2}+{\Delta\over 
\Delta_\theta}\sin^{2}\theta d\phi^{2}\right)~,
\label{eq:string4}
\ee
where $H_1, H_2, \Delta_\theta, \Delta$ are all quadratic functions of  
$r$ given explicitly by
\bea
H_{1} &=& r^{2}+a^{2}\cos^{2}\theta+r(p-2m)+{p\over 
p+q}{(p-2m)(q-2m)\over 2} \nonumber \\ &~&\qquad - {p\over 2m(p+q)} 
\sqrt{(q^{2}-4m^{2})(p^{2}-4m^{2})}~a\cos\theta~,\\
H_{2} &=& r^{2}+a^{2}\cos^{2}\theta+r(q-2m)+{q\over 
p+q}{(p-2m)(q-2m)\over 2} \nonumber \\ &~&\qquad + 
{q\over 2m(p+q)} 
\sqrt{(q^{2}-4m^{2})(p^{2}-4m^{2})}~a\cos\theta~,\\
\Delta_\theta &=& r^{2}-2mr+a^{2}\cos^{2}\theta~,\\
\Delta &=& r^{2}-2mr+a^{2}~,
\eea
 and the 1-forms ${\bf A, B}$ are given by:
\bea
{\bf A} &=& -\left[ Q\left(r+{p-2m\over 2}\right) + 
\sqrt{q^{3}(p^{2}-4m^{2})\over 16m^{2}(p+q)}a\cos\theta\right] H_{2}^{-1} dt
\nonumber \\
&~&\qquad-\left[P(H_{2}+ a^{2}\sin^{2}\theta)\cos\theta+
\sqrt{p(q^{2}-4m^{2})\over 
16m^{2}(p+q)^{3}} \right.
\nonumber \\ &~& 
\left.  \times
\left[(p+q)(pr-m(p-2m))+q(p^{2}-4m^{2})\right]a\sin^{2}\theta 
\right]H_{2}^{-1}d\phi
\label{agauge4} \\
{\bf B} &=& \sqrt{pq}{(pq+4m^{2})r-m(p-2m)(q-2m)\over 2m(p+q)\Delta_\theta}~a 
\sin^{2}\theta d\phi \; .
\label{eq:bgauge4}
\eea
This complicated solution was found by a solution generating technique which uses hidden symmetries of Einstein's equations with Killing fields \cite{Rasheed:1995zv,Matos:1996km,Larsen:1999pp}.

 After dimensional reduction, the five dimensional solution becomes
a four dimensional black hole with the metric:
\be
ds^{2} = - {\Delta_\theta\over\sqrt{H_{1}H_{2}}}(dt + {\bf B})^{2}
+ \sqrt{H_{1}H_{2}}\left( {dr^{2}\over\Delta} + d\theta^{2} + 
{\Delta\over \Delta_\theta}\sin^{2}\theta d\phi^{2}\right)~.
\ee
The matter fields are the gauge field ${\bf A}$ given in 
(\ref{agauge4}), and the dilaton:
\be
e^{-4\phi/\sqrt 3}= {H_{2}\over H_{1}}~.
\ee
The four parameters $m,q,p,a$ appearing in the solution are related to 
the physical parameters $M, Q, P, J$ through:
\bea
G_{4}M&=& {p+q\over 4}
\label{eq:paraM4} \\
G_{4}J &=& {\sqrt{pq}(pq+4m^{2})\over 4(p+q)}~{a\over m}
\label{eq:paraJ4} \\
Q^{2} &=& {q(q^{2}-4m^{2})\over 4(p+q)}
\label{eq:paraQ4} \\
P^{2} &=& {p(p^{2}-4m^{2})\over 4(p+q)}\; .
\label{eq:paraP4}
\eea
The charge parameters $Q$, $P$ were used already in writing the 
solution above. Note that $q,p\geq 2m$, with equality corresponding
to the absence of electric or magnetic charge, respectively. The solutions discussed in the previous section correspond to $a=0$ and either $p=2m$ or $q=2m$.

These rotating black holes have  inner and outer horizons at $\Delta=0$, {\it i.e.}:
\be
r_{\pm} = m \pm \sqrt{m^{2}-a^{2}}~.
\label{eq:horizons4}
\ee
There are two qualitatively different types of extremal limits. \index{extremal limit} Both yield black holes with smooth event horizons. The first is like Kerr: $a=m$. This is called the ``fast rotation" case since $G_4 J > QP$. The other extremal limit is like the charged case discussed earlier. We take $m\to 0$ and $ a\to 0$, but keep the ratio $a/m< 1$ fixed. This is the ``slow rotation" case since $G_4 J < QP$. In this limit,
\be
Q^{2/3} + P^{2/3} = \left({q+p\over 2}\right)^{2/3}
\ee
so the mass of the black hole can be expressed in terms of the charges
\be
2G_4 M = (Q^{2/3} + P^{2/3})^{3/2} \; .
\ee
The striking thing about this formula is that it is independent of the angular momentum $J$! As long as $G_4 J < QP$, one can add angular momentum to an extremal Kaluza-Klein black hole without changing the mass or making it nonextremal. Another surprising aspect about this extremal black hole is that the angular velocity of the horizon vanishes, even though the angular momentum is nonzero. \index{angular velocity} This implies that there is no ergoregion.

The black hole temperature vanishes in both extremal limits. The entropy \index{black hole thermodynamics!entropy}  takes a simple form. For the slowly rotating extremal solution ($G_4 J < QP$), the entropy is
\be
S = 2\pi \sqrt{{Q^2P^2\over G_4^2} - J^2}
\ee
while for the fast rotation case ($G_4 J > QP$), the extremal entropy is
\be
S = 2\pi \sqrt{J^2 - {Q^2P^2\over G_4^2} } \; .
\ee
The only difference is the overall sign inside the square root. When $G_4 J = QP$, the horizon area vanishes and the extremal solution is again singular. The solutions we discussed in the previous section have $J=0$ and only one nonzero charge, so their horizon area  vanishes in the extremal limit as have already discussed.

\section{Inhomogeneous black hole solutions}

For the remainder of this chapter we shall focus on static vacuum five dimensional solutions to Kaluza-Klein theory, but will drop the condition that the metric has an isometry associated to translations about the $S^1$. With the isometry, the only static black hole solution which asymptotes to the Kaluza-Klein vacuum is (\ref{magbs4}) \cite{Mars:2001pz,Gibbons:2002ju}.
We further restrict our attention to solutions with no Kaluza-Klein magnetic charge, so that the only translationally invariant static black hole solution is the black string.
However, we shall see that the full space of static solutions without the circle isometry is very complicated indeed.\footnote{
Much literature exists on this topic, and there are many details and interesting avenues we are not able to cover here. The interested reader is referred to two reviews on this topic \cite{Kol:2004ww,Harmark:2007md}, although we note there has been significant numerical progress since these were written.
}

Without the $S^1$ isometry, the Einstein equations lead to partial differential equations, and it is currently unclear how to find exact solutions. Perturbative techniques give a window onto certain solutions and we shall discuss these shortly. However generally we require numerical methods such as those discussed later in Chapter 10 
to reveal the full and fascinating structure of the space of solutions.

\subsection{Localized black holes}

We will begin our exploration of the general static Kaluza-Klein black hole by taking a limit where we can neglect the fact that the extra dimension is compact. A very small black hole, where by `small' we mean with radius much smaller than the size of the compact dimension $L$, will appear to a nearby observer to look approximately like a 5D Schwarzschild solution. The spherical symmetry of the black hole will be preserved near its horizon, but far away will be broken by the compactification of the spacetime. Conversely the asymptotic $S^1$ translation invariance will be broken by the presence of this black hole which is localized at some particular position on the circle. We term such a solution a `localized' black hole, and these were first discussed in detail by Myers \cite{Myers:1986rx}.

Consider a 5D Schwarzschild solution. We may write this in isotropic coordinates as,
\begin{eqnarray}
\label{eq:Schwarzschild4}
ds^2_{Sch} & =& - \left(  \frac{  \rho_0^2 - 4 \rho^2 }{ \rho_0^2 + 4 \rho ^2 } \right)^2 dt^2 + \left( 1 + \frac{\rho_0^2}{4 \rho ^2} \right)^2 \left( d \rho ^2 + \rho ^2 d \Omega^2_3 \right) \; ,
\end{eqnarray}
where the horizon radius is $\rho_0$, although we should take care that the coordinate position of the horizon, $\rho_h$, is at $\rho_h = \rho_0 / 2$.
Now of course very far from the horizon, for $\rho \gg \rho_h$ this solution becomes flat,
\begin{eqnarray}
\label{eq:SchExp4}
ds^2_{Sch} & = & \left( -  dt^2 +   dr^2 + r^2 d \Omega^2_2 + d y^2 \right) \nonumber \\
&& \quad + \frac{\rho_0^2}{2 \rho ^2} \left( 2 dt^2 +   dr^2 + r^2 d \Omega^2_2 + d y^2 \right) + O(\frac{\rho_0^4}{\rho ^4})
\end{eqnarray}
where we have chosen new coordinates, $r$ and $y$ to write this asymptotic behaviour, where,
\begin{eqnarray}
r = \rho \cos \theta \, , \quad y = \rho \sin \theta \; , 
\end{eqnarray}
so that $\rho ^2 = r^2 + y^2$. The leading term looks just like the Kaluza-Klein vacuum,
\begin{eqnarray}
\label{eq:KKvac4}
ds^2_{vac} & = &  -  dt^2 +   dr^2 + r^2 d \Omega^2_2 + d y^2 \; , 
\end{eqnarray}
except that we cannot identify $y \sim y + L$ as this is not compatible with the subleading term  going as $\sim 1/\rho^2$ in (\ref{eq:SchExp4}). 

Let us try to describe what happens to the metric away from the black hole more precisely. \index{matched asymptotic expansion}
The method we use is based on the `matched asymptotic expansion' approach of \cite{Gorbonos:2004uc,Harmark:2003yz}.
We term the coordinates $r,y$ the `far field' chart. These are adapted to the asymptotic translation symmetry and $y \sim y + L$. The coordinates $\rho,\theta$ form the `near field' chart and these are adapted to the approximate spherical symmetry near the horizon of a small black hole. 
These two charts must overlap. Let us take the extent of the near and far field such that,
\begin{eqnarray}
\mathrm{far\;field:} \; r^2 + y^2 > (1 - \Delta ) \ell^2 \, , \;  \mathrm{near\;field:} \; \rho_h^2 < \rho ^2 < (1 + \Delta) \ell^2
\end{eqnarray}
where $\ell$ is a scale parametrically larger than $\rho_0$ and smaller than $L$ in the limit $\rho_0 / L \to 0$. We might, for example, take $\ell^2 = \rho_0 L$. The constant $\Delta $ is $O(1)$ and ensures the charts overlap so that $0 < \Delta < 1$.

Far from the horizon the black hole appears to be a localized source of mass, and being static, we expect it can be treated as a point mass source in Newtonian perturbation theory. For a point mass source at $r = y = 0$ in the vacuum (\ref{eq:KKvac4}) one may solve the linear problem to obtain the metric in the far field as,
\begin{eqnarray}
ds^2 \simeq ds^2_{vac} + \Phi \left( 2 \,dt^2 + dr^2 + r^2 d \Omega^2_2 + d y^2  \right) \; ,
\end{eqnarray}
where the Newtonian potential is explicitly given as, 
\begin{eqnarray}
\Phi = \frac{4 G_5 M}{3 L r} \frac{ \sinh{ \frac{2 \pi r}{L} } }{ \cosh{ \frac{2 \pi r}{L} } + \cos{ \frac{2 \pi y}{L}  } } \; ,
\end{eqnarray}
where $M$ is the mass as computed in the earlier equation (\ref{kkmass4}) and $\Phi$ is a harmonic function on (\ref{eq:KKvac4}) with a delta function source at the origin \cite{Myers:1986rx}.
In the overlap region then $\rho/L \ll 1$ and we may expand this potential  in the near field coordinates as,
\begin{eqnarray}
\label{eq:farpert4}
\Phi = \frac{4 G_5 M}{3 \pi \rho^2} \left( 1 + \frac{\pi^2}{3} \frac{\rho^2}{L^2} + \frac{ \pi^4 ( 1 - 2 \cos{ 2\theta} ) }{45} \frac{\rho^4}{ L^4} + O\left( \frac{\rho^6}{L^6}\right)  \right) \; ,
\end{eqnarray}
and taking the leading two terms we write the metric in the overlap as,
\begin{eqnarray}
ds^2 \simeq ds^2_{vac} + \frac{4 G_5 M}{3 \pi}  \left( \frac{1}{\rho^2}  + \frac{\pi^2}{3 L^2} \right) \left( 2 \,dt^2 + d\rho^2 + \rho^2 d \Omega^2_3  \right) \; .
\end{eqnarray}
In the near field the matched asymptotic expansion method instructs that we consider perturbations of 5D Schwarzschild. However, for our discussion the perturbation we require can simply be thought of as a global scaling together with a scaling of time. Hence the near field solution remains Schwarzschild which we write as,
\begin{eqnarray}
\label{eq:near4}
ds^2 & =& - (1 + c_t )  \left( \frac{  \rho_0^2 - 4 \rho^2}{  \rho_0^2 + 4 \rho^2 } \right)^2 dt^2 \cr
&+& (1 + c_y ) \left( 1 + \frac{\rho_0^2}{4 \rho^2} \right)^2 \left( d\rho^2 + \rho^2 d \Omega^2_3 \right) \; ,
\end{eqnarray}
where $c_t$ and $c_y$ are constants, with $| c_t |, | c_y | \ll 1$, which give the required perturbative scalings. Then in the overlap, where $\rho \gg \rho_0$ the metric behaves similarly to the expansion (\ref{eq:SchExp4}) giving,
\begin{eqnarray}
ds^2 & \simeq & ds^2_{vac}  +  \frac{\rho_0^2}{2 \rho^2} \left( 2 dt^2 +   d\rho^2 + \rho^2 d \Omega^2_3 \right)  \cr
&-& c_t dt^2 + c_y \left( d\rho^2 + \rho^2 d \Omega^2_3 \right)  \; , 
\end{eqnarray}
We see that in order to `match'  the far and near field asymptotic expansions we must identify,
\begin{eqnarray}
G_5 M = \frac{ 3 \pi } {8} \rho_0^2 \; , \quad - \frac{1}{2} c_t  =  c_y =  \frac{\pi^2 \rho_0^2}{6 L^2} \; .
\end{eqnarray}
We have matched the first two terms in these asymptotic expansions. To proceed further we see that the next term in (\ref{eq:farpert4}) involves an angular dependence and cannot be matched simply by the Schwarzschild metric alone. One must perform static perturbation theory about the Schwarzschild metric, and match this in a multipole expansion. In addition, in the far field one must go beyond Newtonian order. This procedure is detailed in \cite{Gorbonos:2004uc} which shows how to construct the metric in the near and far field order by order in the perturbation parameter $\rho_0 / L$. The simple behaviour we have discussed above arises since the  lowest multipole static perturbations of Schwarzschild (being spherically symmetric) can only be trivial global and time scalings due to Birkhoff's theorem.

Having related the far field parameter, the mass $M$, to the near field parameters $\rho_0$, $c_t$ and $c_y$, we may proceed to compute physical quantities. From the near field solution we may compute the horizon area, $A$, and surface gravity, $\kappa$, perturbatively in powers of $\rho_0 / L$, as, \index{surface gravity}
\begin{eqnarray}
A & \simeq & 2 \pi^2 \rho_0^3 (1 + c_y)^{3/2} = 2 \pi^2 \rho_0^3 \left(1 + \frac{\pi^2 \rho_0^2}{4 L^2} + O\left( \frac{\rho_0^4}{L^4} \right) \right) \; , \nonumber \\
 \kappa & \simeq & \frac{1}{\rho_0} \left( \frac{ 1 + c_t }{ 1 + c_y } \right)^{1/2} = \frac{1}{\rho_0}  \left( 1 - \frac{\pi^2 \rho_0^2 }{4 L^2 } + O\left( \frac{\rho_0^4}{L^4}\right)  \right) \; .
\end{eqnarray}
We know that the mass has leading behaviour $M = \frac{ 3 \pi } {8 G_5} \rho_0^2$. However, we can compute the subleading correction to this using the first law at constant circle size $L$, so we have, $d M = \frac{ \kappa }{ 8 \pi G_5} d A$. Then using the expressions for $A$ and $\kappa$ above yields,
\begin{eqnarray}
M & = & \frac{ 3 \pi } {8 G_5} \rho_0^2 \left(1 + \frac{\pi^2 \rho_0^2}{12 L^2} + O\left( \frac{\rho_0^4}{L^4} \right) \right) \; ,
\end{eqnarray}
and hence we see that for small localized black holes we have the behaviour,
\begin{eqnarray}
\label{eq:smallloc4}
A &=& 32 \left( 2 \pi \right)^{1/2} \left( \frac{ G_5 M }{3} \right)^{3/2} \left( 1 + \frac{\pi G_5 M}{3 L^2} + O\left(  \frac{G_5^2 M^2}{L^4} \right) \right) \; , \nonumber \\
\kappa &=&  \left( \frac{ 8 G_5 M }{3 \pi } \right)^{-1/2} \left( 1 - \frac{5 \pi G_5 M}{9 L^2} + O\left(  \frac{G_5^2 M^2}{L^4} \right) \right) \; ,
\end{eqnarray}
where the leading behaviour is that predicted by Myers \cite{Myers:1986rx}, and the subleading corrections were computed in \cite{Harmark:2003yz,Kol:2003if}.

Another interesting quantity to compute is the proper distance between the poles of the horizon along the axis  of rotational symmetry, $L_{axis}$. One might imagine that as the black hole size is increased there is a corresponding linear response decreasing the length of the axis as $L_{axis} \simeq L \left( 1 - \alpha \frac{\rho_0}{L} \right)$ for some constant $\alpha > 0$. This constant may be computed from our metrics above, and surprisingly one finds it precisely vanishes. There is no linear variation of $L_{axis}$, and this has been termed the `Archimedes effect' \cite{Gorbonos:2004uc}. To leading order, the geometry around the small black hole precisely expands to accommodate it.

Moving beyond perturbation theory we expect that the localized black holes exist as solutions to the full Einstein equations at least for $G_5 M \ll L^2$. Since the 5D Schwarzschild solution is dynamically stable we expect that the localized black hole solutions are similarly dynamically stable in this limit. An interesting question we shall return to later is what happens to this branch of solutions as one increases their mass for fixed $L$. Presumably when $G_5 M \sim L^2$, so that their horizon size becomes of order the asymptotic circle size, they become strongly deformed away from the Schwarzschild geometry. One possibility is that the inter-polar distance $L_{axis}$ actually goes to zero at some finite mass, say $M_\star$, so that the branch of solutions ends. Another possibility is that the Archimedes effect ensures that the geometry always accommodates an increasingly large horizon, and there is no upper mass limit to these solutions. We shall return to this question later, and find evidence that the former suggestion occurs. However, before we discuss this we must introduce a rather exotic static black hole, the inhomogeneous black string.

\subsection{Inhomogeneous black strings} \index{black string!inhomogeneous|(}

Earlier we discussed the simplest static black hole in Kaluza-Klein theory, the black string, which is a straightforward product of the Schwarzschild metric and a circle. Let us write this solution in Schwarzschild form as,
\begin{eqnarray}
\label{eq:ubsmetric4}
ds^2 = - \left( 1 - \frac{r_0}{r} \right) dt^2 + \frac{1}{1 - \frac{r_0}{r}} dr^2 + r^2 d \Omega_2^2 + dy^2 \; .
\end{eqnarray}
We have seen that this solution is dynamically unstable to perturbations with wavenumber $k$ on the circle less than a critical wavenumber $k_c \simeq 0.876 / r_0$. Let us choose our extra dimension to have length $L$ so that $y \sim y  + L$, and take this length to precisely be the critical wavelength for marginal stability, $L = 2 \pi / k_c$. Being marginally stable this black string has an exactly static linear perturbation and 
Gregory and Laflamme realized that this might signal the existence of a new class of static solutions 
\cite{Gregory:1987nb}.
Gubser argued that this deformation does indeed lift to the full non-linear theory to generate an entirely new branch of solutions  \cite{Gubser:2001ac}. Since the perturbation explicitly breaks the translation invariance on the circle these static solutions have the horizon topology of the black string, but are inhomogeneous on the circle. We term them inhomogeneous black strings, in contrast to the solution (\ref{eq:ubsmetric4}) above which we now refer to as a homogeneous black string. Such inhomogeneous black strings were first discussed by Horowitz and Maeda as possible end states of the Gregory-Laflamme (GL) instability \cite{Horowitz:2001cz}. Interestingly, at least for 5D Kaluza-Klein theory it appears that they are all unstable, and therefore cannot serve as such an end state.

Just as for the localized solutions, these inhomogeneous black strings may be constructed perturbatively. For the localized black holes the perturbative limit was when the horizon size was very small compared to $L$. For these inhomogeneous black strings the perturbative limit is when they are very weak deformations of the homogeneous marginally stable solution. We now give an overview of this perturbative construction, essentially following the approach of Gubser \cite{Gubser:2001ac}.
For this discussion we will perform an overall scaling of the solution to fix $r_0 = 1$, so that $L = 2 \pi / 0.876$. We may then write the general metric with the isometries of the inhomogeneous black strings as,
\begin{eqnarray}
ds^2 = - \left( 1 - \frac{1}{r} \right) e^{2 A} dt^2 + e^{2 B} \left( \frac{1}{1 - \frac{1}{r}} dr^2 + dy^2 \right) + r^2 e^{2C} d \Omega_2^2 \; ,
\end{eqnarray}
where the functions $A, B, C$ depend on $r$ and also $y$ due to the inhomogeneity. 
We require that the horizon at $r = 1$ is regular and that asymptotically $A, B, C \rightarrow 0$ which ensures the static Killing vector $\partial / \partial t$ has unit normalization and the circle size is $L$. 

At linear order in perturbation theory we have the marginal GL mode. This takes the form,
\begin{eqnarray}
\label{eq:GLmode4}
A = \lambda \, a(r) \cos{k_c y } \; , \quad B = \lambda  \,b(r) \cos{ k_c y } \; , \quad C = \lambda  \,c(r) \cos{ k_c y } \; ,
\end{eqnarray}
where $\lambda$ is the perturbation parameter, and $a$, $b$, and $c$ are functions of $r$ that must be determined numerically by solving ordinary differential equations.
%
%
Gubser's method gives a systematic way to compute the backreaction of this mode. Non-linear terms in the Einstein equations couple the various Fourier modes on the circle.
The linear term squares to give a source at quadratic order, and since $\cos^2 { k_c y} = \frac{1}{2} ( 1 + \cos{ 2 k_c y})$ one obtains contributions to the backreaction at $O(\lambda^2)$ going as the constant mode and $\cos{2 k_c y}$ on the circle. 
Similarly since $\cos^3 { k_c y}$ decomposes into $\cos{ k_c y}$ and $\cos{ 3 k_c y}$ components,  at cubic order $O(\lambda^3)$ one has backreaction in these Fourier modes, and so on. Thus the full solution generated by the backreaction of the marginal GL mode takes the form,
\begin{eqnarray}
\label{eq:nussoln4}
A = \begin{array}{ccccc} &  \lambda a_{1,1} \cos{ k_c y } \\
+ \lambda^2 a_{2,0} & &  +\lambda^2 a_{2,2}  \cos{ 2 k_c y } \\
&  +\lambda^3 a_{3,1}  \cos{ k_c y } & &  +\lambda^3 a_{3,3}  \cos{ 3 k_c y } \\
+ \lambda^4 a_{4,0} & &  +\lambda^4 a_{4,2}  \cos{ 2 k_c y } & & + \lambda^4 a_{4,2}  \cos{ 4 k_c y } \\
& + \ldots
\end{array}
\end{eqnarray}
where $a_{n,m}$ are functions depending on $r$, and $B$ and $C$ have expansions taking the same form, with coefficient functions $b_{n,m}$ and $c_{n,m}$ respectively. We denote the level $(n,m)$ as being at order $O(\lambda^n)$ and Fourier mode $\cos{ m k_c y}$. The leading linear term is the level $(1,1)$ marginal GL mode, so that $a_{1,1}(r) = a(r)$ in equation (\ref{eq:GLmode4}) and similarly for $b_{1,1}$ and $c_{1,1}$. At the level $(n,m)$ one must solve the problem which we schematically represent as, $\mathcal{L} ( a_{n,m}, b_{n,m}, c_{n,m} ) = \mathcal{S}$
where the left-hand side, $\mathcal{L}$, is a homogeneous linear differential operator acting on the level $(n,m)$ functions $a_{n,m}$, $b_{n,m}$ and $c_{n,m}$, and the right-hand side $\mathcal{S}$ is the source for this inhomogeneous linear problem.  This source originates not only from the component of the $n$-th power of the leading linear term $a_{1,1}$, $b_{1,1}$ and $c_{1,1}$ in the $m$-th Fourier mode, but also from various combinations of the intermediate backreaction orders.

Solving these inhomogeneous linear systems is technical but straightforward, and we will not give the details here.  The level $(n,m)$ linear system can be computed by numerically integrating ordinary differential equations provided all the previous orders $O(\lambda)$, $O(\lambda^2) \ldots O(\lambda^{n-1})$ have been computed. 
In practice computing levels $(1,1)$, $(2,0)$, $(0,2)$ and $(3,1)$ is rather straightforward and that is all we shall require for our discussion. 
Numerical computation gives the following data,
\begin{eqnarray}
k_c = 0.876 \, ; \quad a_{1,1}(1) &=& b_{1,1}(1) = -0.55 \, , \quad c_{1,1}(1) = 1 \nonumber \\
a_{2,0}(1) &=& -0.28 \, , \quad b_{2,0}(1) =0.77 \, , \quad c_{2,0}(1) = 0.80 \nonumber \\
a_{2,2}(1) &=& b_{2,2}(1) = 0.34 \, , \quad c_{2,2}(1) = -0.69 \nonumber \\
a_{3,1}(1) &=& b_{3,1}(1) = -0.24 \, , \quad c_{3,1}(1) = 0 
\end{eqnarray}
which in principle specifies the solution to the levels $(1,1)$, $(2,0)$, $(0,2)$ and $(3,1)$ if one integrates these data out from the horizon to infinity (although one would require more precision than the two significant figures given here).\footnote{
Note that we have chosen the constant circle size `scheme' of \cite{Gubser:2001ac} so that the wavenumber in (\ref{eq:nussoln4}) is unperturbed at higher orders in $\lambda$. 
}
For these levels the functions $a_{n,m}$ , $b_{n,m}$ and $c_{n,m}$ all exponentially decay  asymptotically except for $(2,0)$ which has power law decay going as,
\begin{eqnarray}
b_{2,0} &=& \frac{B_{\infty}}{r}+ O(\frac{1}{r^2})  \, , \quad c_{2,0} = \frac{C_{\infty} \log{r}}{r} + O(\frac{1}{r}) \; , \nonumber \\
 B_{\infty} &=& 0.41 \; , \quad C_{\infty} = - 0.12 \; .
\end{eqnarray}
The variation of the mass $M$, area $A$ and surface gravity $\kappa$ for a non-uniform string takes the form,
\begin{eqnarray}
\label{eq:nubsthermo4}
M & = & M_{GL} \left( 1 + m_2 \lambda^2 + m_4 \lambda^4 + \ldots \right) \; , \quad
A  = A_{GL} \left( 1 + a_2 \lambda^2 + a_4 \lambda^4 + \ldots \right) \; , \nonumber \\
\kappa & = & \kappa_{GL} \left( 1 + \kappa_2 \lambda^2 + \kappa_4 \lambda^4 + \ldots \right) \; ,
\end{eqnarray}
where,
\begin{eqnarray}
M_{GL} = \frac{0.876 L^2 }{ 4 \pi G_5} \; , \quad A_{GL} = \frac{0.876^2 L^3}{\pi} \; , \quad \kappa_{GL} = \frac{\pi}{0.876 L} \; ,
\end{eqnarray}
and we have now rescaled our solution to have a circle size $L$ again. 
Note that the marginal GL perturbation of level $(1,1)$ does not contribute to these at order $O(\lambda)$ as it decays exponentially away from the horizon, and being a harmonic perturbation on the circle, does not change the horizon area at this order. In fact we see the corrections to these quantities only arise from even powers in $\lambda$. 
The quadratic variations are determined from the above data as,
\begin{eqnarray}
\label{eq:values4}
m_2 & = & 3 B_\infty - 2 C_\infty = 1.45 \nonumber \\
a_2 & = & b_{2,0}(1) + 2 c_{2,0}(1) + \frac{1}{4} b_{1,1}(1)^2 + b_{1,1}(1) c_{1,1}(1) + c_{1,1}(1)^2 =  2.90 \nonumber \\
\kappa_2 & = & a_{1,1}(1) - b_{1,1}(1) =  -1.04  
\end{eqnarray}
and thus in a similar manner to the small localized solutions, we have computed the properties of the inhomogeneous string in a perturbative limit.

Let us consider the dynamical stability of the homogeneous black strings. Homogeneous strings with $M < M_{GL}$ are unstable to GL perturbations with wavelength $L$. Indeed for small enough mass, $M < M_{GL}/n$ for integer $n$ the wavelengths $L/2, L/3, \ldots, L/n$ will also be unstable as the higher harmonics of the GL mode fit into the circle. For $M > M_{GL}$ the homogeneous black strings are stable as the unstable GL modes, having minimum wavelength $(M / M_{GL}) L$, cannot fit onto the circle. 

Now let us understand the stability of the weakly inhomogeneous solutions, namely those with $\lambda \ll 1$, so that $M$ is close to $M_{GL}$. We see that since $m_2 > 0$ these inhomogeneous solutions have mass greater than $M_{GL}$ and hence coexist at the same mass as a stable homogeneous black string. Let us compare the areas of these two solutions.
Consider moving an infinitesimal distance $d\lambda$ along the inhomogeneous black string branch starting at $\lambda = 0$.
Then the first law, $dM = \kappa \, dA / (8 \pi G_5)$, for fixed circle size implies at order $O(\lambda^2)$ that $2 m_2 = a_2$, which we see is consistent with the numerical data above. At order $O(\lambda^4)$ it implies that $2 m_4 - a_4 - \frac{1}{2} \kappa_2 a_2 = 0$. A homogeneous string has area $A_{h} = \left( \frac{16 \pi G_5^2}{L} \right) M^2$ for a mass $M$, and so we can compute the fractional area difference between the inhomogeneous and homogeneous strings for a mass $M$ as, 
\begin{eqnarray}
\label{eq:entropydiff4}
\frac{\Delta A}{A_h} = \frac{A - A_h}{A_h} &=&  \left( a_2 - 2 m_2 \right) \lambda^2 + \left( a_4 - 2 a_2 m_2 + 3 m_2^2 - 2 m_4 \right) \lambda^4 + O(\lambda^6) \nonumber \\
& = & - m_2 \left( m_2 + \kappa_2 \right) \lambda^4 + O(\lambda^6) \simeq - 0.59 \lambda^4
\end{eqnarray}
where the quadratic term vanishes by the first law, and the quartic term is precisely determined by the data we have given above. 
We arrive at the result that at fixed mass and fixed circle size, the weakly inhomogeneous string has a lower area than the stable homogeneous string of the same mass. 
Since area cannot decrease in a dynamical process this argument shows that the weakly inhomogeneous strings are dynamically unstable to perturbations that preserve their mass, and deform them into the homogeneous strings.\index{instability!black string} It has currently not been demonstrated whether this perturbative instability can be seen in linear perturbation theory or only at higher orders. However, we believe it is likely to manifest as a linear dynamical instability, and we shall later assume this is the case.
A reasonable supposition is that such a linear instability generates an evolution that ends at a stable homogeneous black string with similar mass.

As with the localized solutions, an interesting question is what happens to the branch of inhomogeneous solutions as one deforms past the weakly inhomogeneous regime so that the perturbative approach above breaks down. In the next section we shall consider this.\index{black string!inhomogeneous|)}

\subsection{The space of static black hole solutions}

We have seen that the homogeneous black strings exist for all mass, with fixed circle size, but are stable only for sufficiently large mass, $M > M_{GL} \sim O( L^2 / G_5 )$. At low masses, $M \ll L^2 / G_5$, we have another branch of solutions, the localized black holes, which we expect to be dynamically stable. In the intermediate mass range for masses just above $M_{GL}$ we also have the inhomogeneous black strings, which we have argued are unstable. We already see there is no uniqueness for static solutions to Kaluza-Klein theory for fixed mass and circle size.

We have posed the question of what happens to localized black holes as they become large, and also what happens to the inhomogeneous black strings as they become increasingly inhomogeneous. 
Harmark and Obers \cite{Harmark:2002tr} argued that the localized black holes might continuously connect to the inhomogeneous black string solutions proposed by Horowitz and Maeda. Kol used a simple Morse theory argument to deduce that the simplest scenario is for the localized black holes to continuously connect to the inhomogeneous black string branch found by Gubser \cite{Kol:2002xz}. Furthermore he predicted that the two branches connect via a singular solution that mediates a topology change of the horizon. \index{horizon topology|(}
This is analogous to the topology change between hyperbolic and parabolic conic sections, which is mediated by the singular  section taken through the apex of the cone. Indeed there is a static, spherically symmetric Ricci flat conical geometry,
\begin{eqnarray}
ds^2_{cone} = d \alpha^2 + \frac{1}{3} \alpha^2 \left( d \beta^2 - \sin^2{\beta} dt^2 \right) + \frac{1}{3} \alpha^2 d\Omega^2_2 \; ,
\end{eqnarray}
which is singular at its apex $\alpha = 0$. The base of the cone is the product of the 2-sphere with the 2D de Sitter space $(d \beta^2 - \sin^2{\beta} dt^2)$, with $\beta \in [ 0 , 2 \pi ]$ where $\beta = 0$ and $\beta = \pi$ are Killing horizons with respect to $\partial / \partial t$. 
These horizons are connected (although not smoothly) at the apex of the cone where the de Sitter factor is zero sized.
Kol proposed that this gives a local model for the geometry of the topology changing solution at the singular point of the horizon. Moving away from the singular topology changing solution to the inhomogeneous black string branch would resolve the sphere so that it was finite in size at $\alpha = 0$, in accord with the inhomogeneous strings having no exposed axis of symmetry, and the two components of the horizon with $\beta = 0$ and $\beta =  \pi$ would be smoothly connected.
Conversely, moving to the localized black hole branch would resolve the de Sitter factor, so that the horizons at $\beta = 0$ and $\beta =  \pi$ would not touch, and $\alpha = 0$ would be the exposed axis of symmetry.

In order to make progress in understanding the extension of the localized black hole and inhomogeneous black string branches we must resort to numerical work. The inhomogeneous black strings were first computed in \cite{Wiseman:2002zc} and the localized black holes in \cite{Kudoh:2003ki,Sorkin:2003ka,Kudoh:2004hs}. Recently a standard approach to numerically construct static vacuum solutions has been developed \cite{Headrick:2009pv} and will be discussed in Chapter 10. Unlike the previous methods employed, for example \cite{Wiseman:2002zc}, this approach is geometrically elegant being covariant and may be applied to problems depending non-trivially on as many coordinates as one likes. In practice these new methods work very well. The results presented here for both inhomogeneous black strings and localized black holes are those computed in \cite{Headrick:2009pv}. 

As we shall now see, the results are compatible with Kol's prediction of a topology changing merger of the localized black holes and inhomogeneous black strings discussed above. \index{horizon topology|)} The numerical calculations have not yet been adapted to explore the singular potential merger point, and inevitably break down for both the localized and inhomogeneous string branches as this point is approached. 
We note that it is still early days for these methods, and we fully expect that future work in coming years will significantly improve on the solutions reproduced here.

\begin{figure}[h]
\makebox[\textwidth]{
\includegraphics[width=15.5cm]{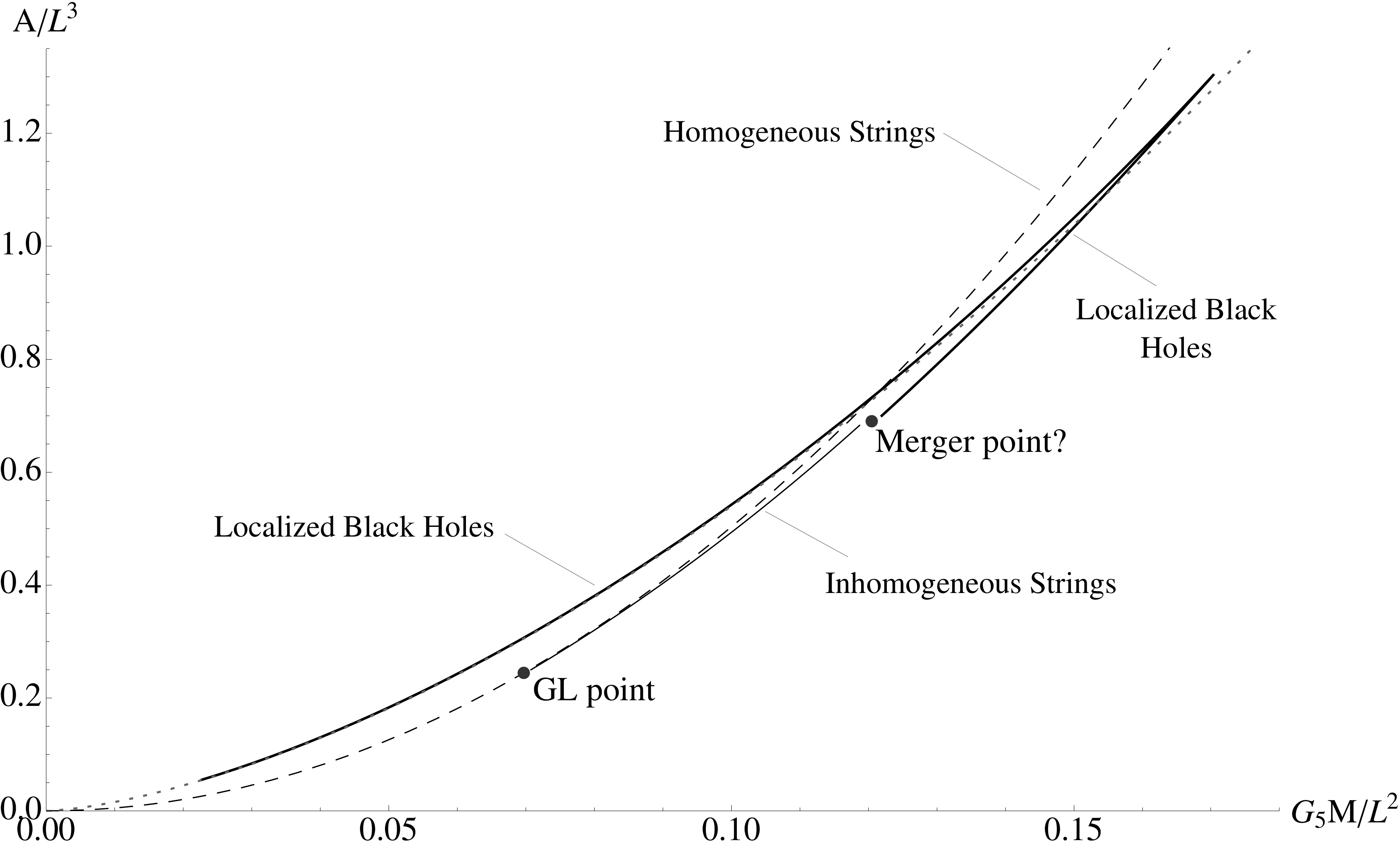} 
}
\caption{Plot of area $A$ against mass $M$ for fixed Kaluza-Klein circle size $L$  for the homogeneous (dashed) and inhomogeneous (solid thin line) black strings and the localized black holes (solid thick line). The curves for the latter two are the numerical solutions of  \cite{Headrick:2009pv}. The inhomogeneous black strings and localized black holes are compatible with a merger at $G_5 M_{\star} / L^2 \simeq 0.12$, and there is a maximum mass localized solution, with $G_5 M_{max} / L^2 \simeq 0.17$. The gray dotted line is the small localized black hole approximation in equation (\ref{eq:smallloc4}) and we see the approximation is excellent for increasing mass up to $M \sim M_{max}$. 
}
\label{fig:AvsM}
\end{figure}

\begin{figure}[h]
\makebox[\textwidth]{
\includegraphics[width=15.5cm]{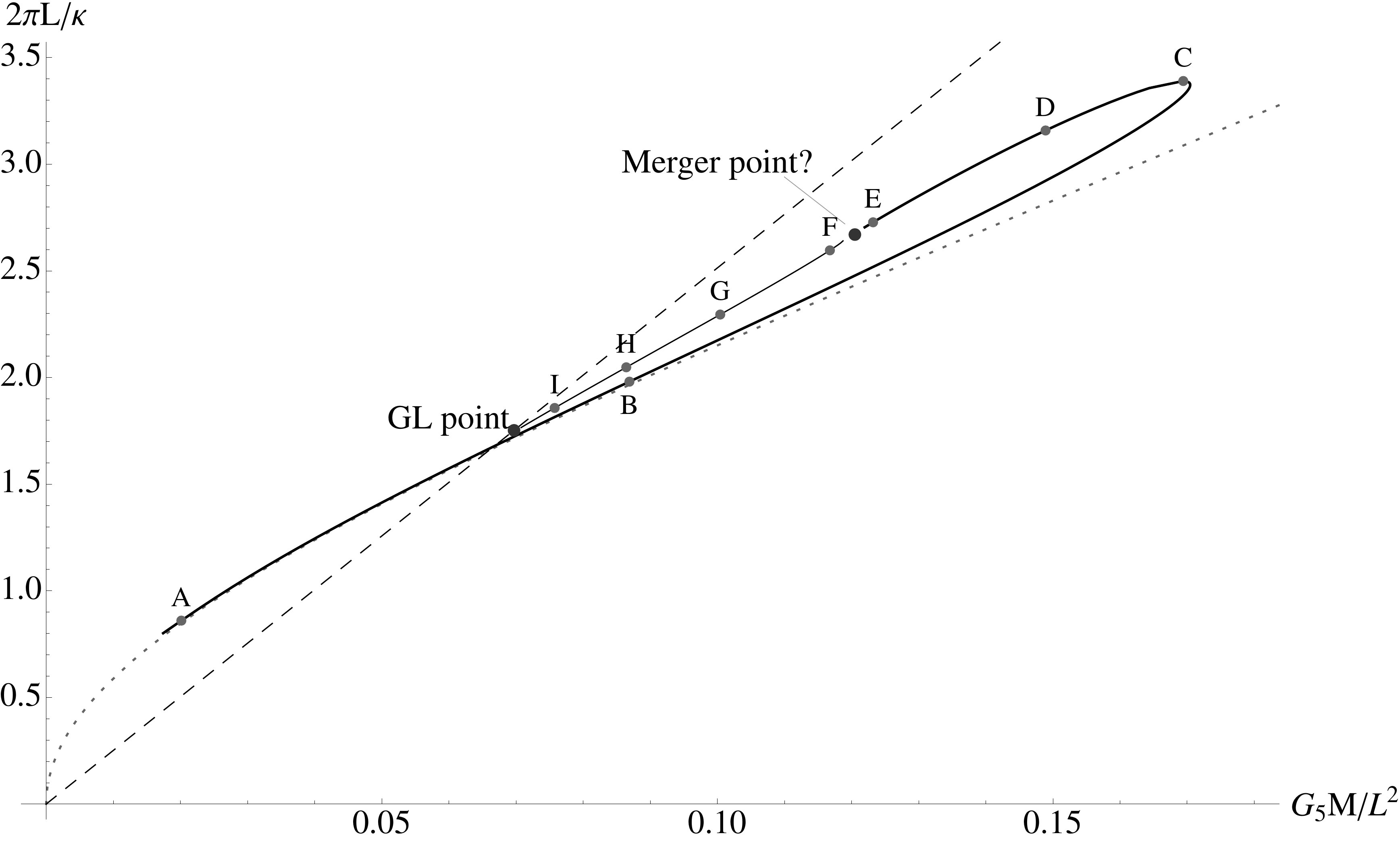} 
}
\caption{ Plot of inverse temperature $2\pi/\kappa$ against mass $M$ for fixed circle size $L$ for the same solutions as in the previous figure.
The labels `A' to `I' denote solutions whose horizon embeddings are displayed in the following figure. Again we see consistency with a merger. We also note that there is a minimum surface gravity solution at $M_{\kappa}$ (near the label `C') which is rather close to, but slightly less than $M_{max}$. The perturbative approximation in equation (\ref{eq:smallloc4}) is also plotted and again very good agreement is seen for localized solutions with increasing mass up to around $M \sim M_{max}$.
}
\label{fig:TvsM}
\end{figure}

Let us begin in Figs. \ref{fig:AvsM} and \ref{fig:TvsM} by showing the curves of area and surface gravity, plotted as the inverse temperature $1/T = 2 \pi / \kappa$, against mass for fixed circle size $L$. Quantities are made dimensionless with appropriate powers of the circle size $L$.
We find a very interesting behaviour for the localized black holes, namely that there is a maximum mass solution with $G_5 M_{max} \simeq 0.17 L^2$.  We note that since $dM = \kappa dA / 8 \pi G_5$ for fixed circle size $L$, then any extremum of $M$ will similarly be an extremum of $A$. Hence the area against mass curve has a cuspy profile at $M = M_{max}$. We also see from Fig. \ref{fig:TvsM} that there is a minimum surface gravity (or equivalently temperature) solution, at $M = M_{\kappa}$. The actual value of $M_\kappa$ is numerically close to that of $M_{max}$, as is clear visually, but we emphasize that since the curve of $1/\kappa$ against $M$ is smooth for the localized solutions about the maximum mass, as we expect and indeed see numerically, the point of maximum mass and minimum surface gravity cannot coincide, and $M_\kappa < M_{max}$.
This leads to the peculiar conclusion that the localized solutions in the narrow range $M_ \kappa < M < M_{max}$ actually have positive specific heat.

We confirm that the perturbative prediction computed above in equation (\ref{eq:smallloc4}) provides a good description of the localized behaviour for small mass. In both figures this approximation is plotted, and we see surprisingly good agreement for increasing mass up to $M_{GL}$ and even past this point to around $M \sim M_{max}$. These approximations have no maximum mass, and hence cannot agree past this point in the localized branch, and we expect non-perturbative effects to become important there. 

The inhomogeneous black string branch departs from the homogeneous one at the GL point $M = M_{GL}$. Moving away from this point the deformation of the horizon becomes greater. The degree of inhomogeneity, measured by the ratio of the maximum and minimum 2-sphere radii of the horizon, increases monotonically. The most striking feature of these plots is that moving as far along the localized and inhomogeneous branches as possible, both branches appear to be compatible with joining at some mass $M_{\star}$ with $G_5 M_\star \simeq 0.12 L^2$. This is an indication that indeed these branches both end at the same singular topology changing solution. Other physical quantities that we have not displayed here also support this conclusion.

\begin{figure}[h]
\makebox[\textwidth]{
\includegraphics[width=14.5cm]{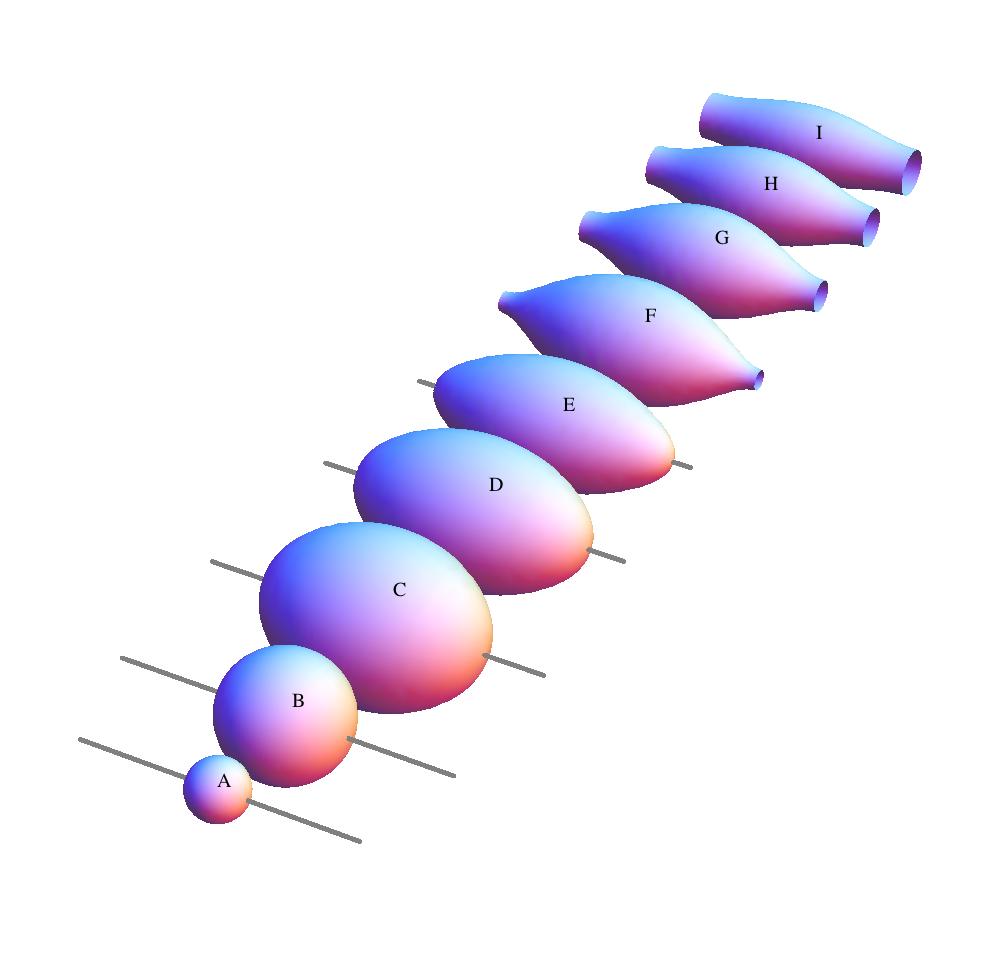}
}
\caption{
Figure showing the spatial geometry of the horizon for a number of localized black holes and inhomogeneous black strings all with the same asymptotic circle size. The embeddings are labelled `A' to `I' and correspond to the solutions annotated in the previous figure. We emphasize that these geometries are those actually found in the numerical solutions of \cite{Headrick:2009pv}. For the localized black holes the proper length of the axis of symmetry is also depicted.
}
\label{fig:embedding}
\end{figure}

In Fig. \ref{fig:embedding} we show several embeddings of the horizon for various localized and inhomogeneous string solutions. We see graphical confirmation that the most inhomogeneous strings found appear to be close to having the minimum 2-sphere of the horizon go to zero size. The localized black holes also appear to be close to the poles meeting as the axis of symmetry shrinks to zero length. Near the potential merger, the geometries appear to be similar away from the singular region.

We note that by uncompactifying the solutions above, and recompactifying with circle size an integer multiple of $L$ one can construct multi-centered localized black hole solutions although all such solutions presumably would be unstable.

The metric for the inhomogeneous black strings has been tested to see explicitly that the cone geometry advocated by Kol emerges for solutions with $M \simeq M_{\star}$ in the region surrounding the minimum radius of the horizon. 
Strings in various dimensions were found, and the appropriate cone geometries were indeed seen to emerge \cite{Sorkin:2006wp}. Similar tests have not yet been performed on the localized solutions, but we believe it is very plausible that this cone geometry will again be seen.

\subsection{Stability of the static black hole solutions}

Having seen an elegant proposal for the space of static solutions and numerical evidence supporting it we now discuss their likely dynamical stability. From the earlier Fig. \ref{fig:AvsM} we see that for fixed circle size the localized solutions dominate the area for a given mass when $M < M_{crit}$ where $G_5 M_{crit} \simeq 0.12 L^2$, and for $M > M_{crit}$ the homogeneous black strings dominate the area. Thus at fixed mass (the `microcanonical ensemble') we expect that the globally stable solution is a localized black hole for $M < M_{crit}$ and a homogeneous string for $M > M_{crit}$.

This does not tell us about local dynamical stability of the solutions. For example, the homogeneous strings with $M_{GL} < M < M_{crit}$ are dynamically stable to small perturbations even though they do not globally dominate the area, and can be deformed to a more stable solution with the same mass for a sufficiently large deformation.

Let us now consider this local stability. We have argued that the small localized solutions are perturbatively stable and the weakly inhomogeneous black strings are perturbatively unstable.  Furthermore it appears that these two branches should be thought of as one, joined together by a singular topology changing solution. If these assertions are correct the simplest situation is that either i) the entire inhomogeneous branch and some of the localized branch are unstable to a single perturbation mode, or alternatively ii) the instability ends at some point along the inhomogeneous branch and so there are strongly deformed stable inhomogeneous solutions and all the localized black holes are stable. For either of these scenarios to be true the instability should smoothly turn off as one moves in the space of solutions, and hence there will be a localized solution (for scenario i) or an inhomogeneous string (scenario ii) with a marginal static perturbation.

The localized branch has precisely such a candidate marginal static perturbation as we shall now argue, indicating that scenario i) is realized. \index{instability!spherical black hole} Since we are considering the microcanonical ensemble we require a static perturbation that leaves the mass invariant. (A static perturbation that changes the mass represents a change in the black hole, not a perturbation of a given black hole.) Take the perturbation of the metric for a localized solution to be that generated by moving infinitessimally along the branch of solutions (i.e. the perturbation is the tangent to the space of solutions there). For a generic solution such a perturbation will change the mass to linear order. However precisely at the point $M = M_{max}$ the perturbation will leave the mass invariant to linear order. This then gives the required candidate static perturbation that is both regular at the horizon and leaves the mass invariant.\footnote{We thank Harvey Reall for an important discussion clarifying this issue.}
Physically we may gain some intuition for this change in stability at the maximum mass solution by considering taking a small (presumably stable) localized black hole and gradually dropping matter into it. Initially one may expect the mass will grow as it absorbs the matter. However, when it reaches the localized maximum mass solution with $M = M_{max}$ (and correspondingly maximal area) it can no longer remain a stable solution. Adding more matter will force the area to increase during the ensuing dynamics implying the solution cannot settle back to any solution near $M = M_{max}$ since these all have lower area.

Let us summarize our stability discussion. The numerical solutions imply that localized black holes with $M < M_{crit}$ are globally dynamically stable, while homogeneous strings with $M > M_{crit}$ are globally stable. 
We have argued that the simplest picture of linear stability is that the localized black holes are linearly stable for small mass, and remain stable as their mass is increased up to the value $M = M_{max}$. Moving further along this localized branch these solutions become unstable to a single perturbation mode, and this instability continues through the topology change, and all the way along the inhomogeneous black string branch. A further comment is that whilst dynamically unstable solutions may be interesting in terms of understanding the moduli space of static vacuum solutions, presumably they play no physical role. Whilst we have had to resort to sophisticated numerical work to elucidate the full structure of solutions, from a physical perspective the interesting (i.e. stable) solutions are the simplest ones. The homogeneous black strings we know analytically, and the stable localized solutions with $M < M_{max}$ appear to be rather well described by the perturbative construction we gave earlier.

Sorkin has shown this picture  changes remarkably in higher dimensions \cite{Sorkin:2004qq}. Treating the dimension as a continuous parameter $D$, again with a single compact Kaluza-Klein circle, so that previously in this chapter we have considered only $D = 5$, then one finds that the sign of $m_2$ in equation (\ref{eq:nubsthermo4}), and correspondingly that of the area difference in equation (\ref{eq:entropydiff4}), changes above a critical dimension $D_\star \simeq 13.5$. 
Now weakly inhomogeneous strings have masses $M < M_{GL}$, and are expected to be stable and the end state of the GL instability of homogeneous strings with mass  just below $M_{GL}$.
Assuming the localized solutions and inhomogeneous strings still merge at some mass $M_\star$, then the simplest picture is that for $M < M_\star$ the localized black holes exist and dominate the area for fixed mass over the homogeneous strings, and for $M_\star < M < M_{GL}$ the inhomogeneous strings exist and dominate the homogeneous ones.
An interesting question is whether there are inhomogeneous black strings that are stable for $D < D_\star$. We have argued for $D = 5$ they are likely to all be unstable. However, for $D$ close to, but less than $D_\star$, continuity implies the strongly inhomogeneous black strings will remain stable, even though the weakly inhomogeneous ones will become unstable. Whether stability persists for strongly inhomogeneous black strings down to $D = 10$ or $D = 11$, the maximum dimensions proposed in quantum theories of gravity, is currently unclear. 

We conclude this discussion with the caveat that the picture advocated here is the simplest one compatible with the numerical data currently available. We may learn that the situation is considerably more complicated than we now expect, and it is certainly possible that there are even more exotic static  black hole solutions waiting to be discovered.

\end{document}